%% file: sigir2018-pytrec_eval.tex
\pdfoutput=1

\documentclass[sigconf,natbib,9pt]{acmart}

\copyrightyear{2018} 
\acmYear{2018} 
\setcopyright{acmlicensed}
\acmConference[SIGIR '18]{The 41st International ACM SIGIR Conference on Research \& Development in Information Retrieval}{July 8--12, 2018}{Ann Arbor, MI, USA}
\acmBooktitle{SIGIR '18: The 41st International ACM SIGIR Conference on Research \& Development in Information Retrieval, July 8--12, 2018, Ann Arbor, MI, USA}
\acmPrice{15.00}
\acmDOI{10.1145/3209978.3210065}
\acmISBN{978-1-4503-5657-2/18/07}

\usepackage[subtle, wordspacing=normal, tracking=normal, bibnotes, charwidths=normal, leading, indent, lists, paragraphs=normal, mathspacing=normal, bibliography=normal]{savetrees}
\usepackage{amsmath}
\usepackage{booktabs}
\usepackage{color}
\usepackage{dsfont}
\usepackage{enumitem}
\usepackage{etoolbox}
\usepackage{footnote}
\usepackage{hyperref}
\usepackage{grffile}
\usepackage{mathtools}

\usepackage{mfirstuc}
\usepackage[frozencache]{minted}
\usepackage{multirow}
\usepackage[np, autolanguage]{numprint}
\usepackage[nodisplayskipstretch]{setspace}
\usepackage{subcaption}
\usepackage{tabularx}
\usepackage{tikz}
\usepackage{tikz-3dplot}
\usepackage{todonotes}
\usepackage{paralist}
\usepackage[labelfont=bf,textfont=bf,skip=0pt]{caption}
\usepackage[normalem]{ulem}

\AtBeginEnvironment{minted}{\singlespacing%
    \fontsize{8}{8}\selectfont}

\setcitestyle{square,comma,numbers,sort&compress,sectionbib}

\newminted{python}{mathescape, numbersep=5pt, autogobble, framesep=2mm, fontsize=\footnotesize}

\newfloat{algorithm}{t}{lop}
\floatname{algorithm}{Code snippet}

\newcommand{\TRECEval}{\texttt{trec\_eval}}
\newcommand{\PyTRECEval}{\texttt{pytrec\_eval}}
\newcommand{\NDCG}{NDCG}
\newcommand{\MAP}{MAP}

\newcommand{\PyTRECEvalURL}{\url{https://github.com/cvangysel/pytrec_eval}}

\newcommand{\FirstRQ}{What speedup do we obtain when using \PyTRECEval{} over \TRECEval{} (serialize-invoke-parse workflow)?}
\newcommand{\SecondRQ}{How fast is \PyTRECEval{} compared to native Python implementations of IR evaluation measures?}

\allowdisplaybreaks

\begin{document}

\title{Pytrec\_eval: An Extremely Fast Python Interface to trec\_eval}
\titlenote{Open-source implementation is available at \PyTRECEvalURL{}.}

\author{Christophe Van Gysel}
\orcid{0000-0003-3433-7317}
\affiliation{%
\institution{University of Amsterdam}
\city{Amsterdam}
\country{The Netherlands}
}
\email{chris@stophr.be}

\author{Maarten de Rijke}
\orcid{0000-0002-1086-0202}
\affiliation{%
\institution{University of Amsterdam}
\city{Amsterdam}
\country{The Netherlands}
}
\email{derijke@uva.nl}

\begin{abstract}
We introduce \PyTRECEval{}, a Python interface to the \TRECEval{} information retrieval evaluation toolkit. \PyTRECEval{} exposes the reference implementations of \TRECEval{} within Python as a native extension. We show that \PyTRECEval{} is around one order of magnitude faster than invoking \TRECEval{} as a sub process from within Python. Compared to a native Python implementation of \NDCG{}, \PyTRECEval{} is twice as fast for practically-sized rankings. Finally, we demonstrate its effectiveness in an application where \PyTRECEval{} is combined with Pyndri and the OpenAI Gym where query expansion is learned using Q-learning.
\end{abstract}



\maketitle

\input{introduction}
\input{toolkit}
\input{results}
\input{application}
\input{conclusions}

\smallskip\noindent
\textbf{Acknowledgements.}
This research was supported by
Ahold Delhaize,
Amsterdam Data Science,
the Bloomberg Research Grant program,
the China Scholarship Council,
the Criteo Faculty Research Award program,
Elsevier,
the European Community's Seventh Framework Programme (FP7/2007-2013) under
grant agreement nr 312827 (VOX-Pol),
the Google Faculty Research Awards program,
the Microsoft Research Ph.D.\ program,
the Netherlands Institute for Sound and Vision,
the Netherlands Organisation for Scientific Research (NWO)
under pro\-ject nrs
CI-14-25, 
652.\-002.\-001, 
612.\-001.\-551, 
652.\-001.\-003, 
and
Yandex.
All content represents the opinion of the authors, which is not necessarily shared or endorsed by their respective employers and/or sponsors.

\bibliographystyle{abbrvnatnourl}
\bibliography{sigir2018-pytrec_eval}

\end{document}

%% file: introduction.tex

\section{Introduction}
\label{sec:introduction}

Evaluation is a crucial component of any information retrieval (IR) system \citep{Harman2011information}. Reusable test collections and off-line evaluation measures \citep{Sanderson2010testcollections} have been the dominating paradigm for experimentally validating IR research for the last 30 years. 
The popularity and ubiquity of off-line IR evaluation measures is partly due to the Text REtrieval Conference (TREC) \citep{TREC}. 
TREC led to the development of the \TRECEval{}\footnote{\url{https://github.com/usnistgov/trec_eval}} software package that is the standard tool for evaluating a collection of rankings. 
The \TRECEval{} tool allows IR researchers to easily compute a large number of evaluation measures using standardized input and output formats. 
For a document collection, a test collection of queries with query/document relevance information (i.e., \texttt{qrel}) and a set of rankings generated by a particular IR system (i.e., a system \texttt{run}) for the test collection queries, \TRECEval{} outputs a standardized output format containing evaluation measure values. The adoption of \TRECEval{} as an integral part of IR research has led to the following benefits:
\begin{inparaenum}[(a)]
	\item \emph{standardized formats} for system rankings and query relevance information such that different research groups can exchange experimental results with minimal communication, and
	\item \emph{open-source reference implementations of evaluation measures}---provided by a third party (i.e., NIST)---that promotes transparent and consistent evaluation.
\end{inparaenum}

While the availability of \TRECEval{} has brought many benefits to the IR community, it has the downside that it is available only as a standalone executable that is interfaced by passing files with rankings and ground truth information. 
In recent years, the Python programming language has risen in popularity due to its feature richness (i.e., scientific libraries and data structures) and holistic language design \citep{Koepke2010}. 
Research progresses at a rate proportional to the time it takes to implement an idea, and consequently, scripting languages (e.g., Python) are preferred over conventional programming languages \citep{Prechelt2000}. 
Within IR research, retrieval systems are often implemented and optimized using Python (e.g., \citep{VanGysel2017pyndri,Li2018bayesian}) and for their evaluation \TRECEval{} is used. However, invoking \TRECEval{} from Python is expensive as it involves
\begin{inparaenum}[(1)]
	\item serializing the internal ranking structures to disk files,
	\item invoking \TRECEval{} through the operating system, and
	\item parsing the \TRECEval{} evaluation output from the standard output stream.
\end{inparaenum}
This workflow is unnecessarily inefficient as it incurs
\begin{inparaenum}[(a)]
	\item a double I/O cost when the ranking is first serialized by the Python script and subsequently parsed by \TRECEval{}, and
	\item a context-switching overhead as the invocation of \TRECEval{} needs to be processed by the operating system.
\end{inparaenum}

We introduce \PyTRECEval{} to counter these excessive efficiency costs and avoid a wild growth of ad-hoc Python-based evaluation measure implementations. \PyTRECEval{} builds upon the \TRECEval{} source code and exposes a Python-first interface to the \TRECEval{} evaluation toolkit as a native Python extension. Rankings constructed in Python can directly be passed to the evaluation procedure, without incurring disk I/O costs; evaluation is performed using the original \TRECEval{} implementation. Due to \PyTRECEval{}'s implementation as a native Python extension, context-switching overheads are avoided as the evaluation procedure and its invocation reside within the same process.
Next to improved efficiency, \PyTRECEval{} brings the following benefits:
\begin{inparaenum}[(a)]
	\item current and future reference \TRECEval{} implementations of IR evaluation measures are available within Python, and
	\item as the evaluation measures are implemented in C, their execution are typically faster than native Python-based alternatives.
\end{inparaenum}
\newcommand{\RQItem}[2]{\textbf{(RQ#1)} #2}
The main purpose of this paper is to describe \PyTRECEval{}, provide empirical evidence of the speedup that \PyTRECEval{} delivers, and showcase the use of \PyTRECEval{} in a reinforcement learning application. We ask the following questions:
\RQItem{1}{\FirstRQ{}}
\RQItem{2}{\SecondRQ{}}
We also present a demo application that combines Pyndri \citep{VanGysel2017pyndri} and \PyTRECEval{} in a query formulation reinforcement learning setting and provide the environment and the reward signal, integrated within the OpenAI Gym \citep{OpenAI2016}.

%% file: toolkit.tex

\section{Evaluating using Pytrec\_eval}

The \PyTRECEval{} library has a minimalistic design. Its main interface is the \texttt{RelevanceEvaluator} class. The \texttt{RelevanceEvaluator} class takes as arguments
\begin{inparaenum}[(1)]
	\item query relevance ground truth, a dictionary of query identifiers to a dictionary of document identifiers and their integral relevance level, and
	\item a set of evaluation measures to compute (e.g., \texttt{ndcg}, \texttt{map}).
\end{inparaenum}
\begin{algorithm}
\begin{minted}[frame=lines]{python}
qrel = {'q1': {'d1': 1, 'd2': 0},
        'q2': {'d2': 1}}
run = {'q1': {'d1': 0.5, 'd2': 2.0},
       'q2': {'d1': 0.5, 'd2': 0.6}}

evaluator = pytrec_eval.RelevanceEvaluator(
	qrel, {'map', 'ndcg'})
result = evaluator.evaluate(run)

# result equals
# {'q1': {'map': 0.5, 'ndcg': 0.6309297535714575},
#  'q2': {'map': 1.0, 'ndcg': 1.0}}
\end{minted}
\caption{Minimal example of how \PyTRECEval{} can be used to compute IR evaluation measures. Evaluation measures (\NDCG{}, \MAP{}) are computed for two queries---\texttt{q1} and \texttt{q2}---and two documents---\texttt{d1} and \texttt{d2}---where for \texttt{q2} we only have partial relevance (\texttt{d1} is assumed to be non-relevant).\label{example:main}}
\end{algorithm}
Code snippet~\ref{example:main} shows a minimal example on how \PyTRECEval{} can be used to evaluate a ranking. Rankings are encoded by a mapping from document identifiers to their retrieval scores. Internally, \PyTRECEval{} sorts the documents in decreasing order of retrieval score. This behavior mimics the implementation of \TRECEval{}, which ignores the order of documents within the user-provided file, and only considers the document scores. Similar to \TRECEval{}, document ties, which occur when two documents are assigned the same score, are broken by secondarily sorting on document identifier. Query relevance ground truth is passed to \PyTRECEval{} in a similar way to document scores, where relevance is encoded as an integer rather than a floating point value.

Beyond measures computed over the full ranking of documents, \PyTRECEval{} also supports measures computed up to a particular rank $k$. The values of $k$ are the same as the ones used by \TRECEval{}. For example, measures \texttt{ndcg\_cut} and \texttt{P} correspond to \NDCG{}@$k$ and precision@$k$, respectively, with $k = 5, 10, 15, 20, 30, 100, 200$, $500$, $1000$. The set of supported evaluation measures is stored in the \texttt{pytrec\_eval.}\texttt{supported\_measures} property and the identifiers are the same as used by \TRECEval{} (i.e., running \TRECEval{} with arguments \texttt{-m ndcg\_cut --help} will show documentation for the \NDCG{}@$k$ measure). To mimic the behavior of \TRECEval{} to compute all known evaluation measures (i.e., passing argument \texttt{-m all\_trec} to \TRECEval{}), just instantiate \texttt{RelevanceEvaluator} with \texttt{pytrec\_eval.supported\_measures} as the second argument.

%% file: results.tex

\section{Benchmark results}

As demonstrated above, \PyTRECEval{} conveniently exposes popular IR evaluation measures within Python. However, the same functionality could be exposed by invoking \TRECEval{} in a serialize-invoke-parse workflow---or---by implementing the evaluation measure natively in Python. In this section we provide empirical benchmark results that show that \PyTRECEval{}, beyond its convenience, is also faster at computing evaluation measures than these two alternatives (i.e., invoking \TRECEval{} or native Python).

\smallskip
\noindent
\textbf{Experimental setup.} For every hyperparameter configuration, the runtime measurement was repeated 20 times and the average runtime is reported. Speedup denotes the ratio of the runtime of the alternative method (i.e., \TRECEval{} or native Python) over the runtime of \PyTRECEval{} and consequently, a speedup of $1.0$ means that both methods are equally fast. When invoking \TRECEval{} using the serialize-invoke-parse workflow, rankings are written from Python to storage without sorting, as \TRECEval{} itself sorts the rankings internally. The resulting evaluation output is read from \texttt{stdout} to a Python string and we do not extract the measure values, as different parsing strategies can lead to large variance in runtime. For the native Python implementation, we experimented with different open-source implementations of the \NDCG{} measure and adapted the fastest implementation as our baseline. The implementation does not make use of NumPy or other scientific Python libraries as
\begin{inparaenum}[(a)]
	\item we wish to compare to native Python directly and
	\item the NumPy-based implementations we experimented with were less efficient than the native implementation we settled with, as NumPy-based implementations require that the rankings are encoded in dense arrays before computing evaluation measures.
\end{inparaenum}
The evaluated rankings and ground-truth were synthesized by assigning every document a distinct ranking score in $\mathbb{N}$ and a relevance level of $1$. This allows us to evaluate different evaluation measure implementations with rankings and query sets of different sizes.
Experiments were run using a single Intel Xeon CPU (E5-2630 v3) clocked at \numprint{2.4}GHz, DDR4 RAM clocked at \numprint{2.4}GHz, an Intel SSD (DC S3610) with sequential read/write speeds of 550MB/s and 450MB/s, respectively, and a hard disk drive (Seagate ST2000NX0253) with a rotational speed of 7200 rpm.
All code used to run our experiments is available under the MIT open-source license.\footnote{The benchmark code can be found in the \texttt{benchmarks} sub-directory of the \PyTRECEval{} repository; see the footnote on the first page.}

\newcommand{\RQRef}[1]{\textbf{RQ#1}}
\newcommand{\RQAnswer}[3]{
\begin{description}
	\item[\RQRef{#1}] #2
\end{description}
\noindent #3
}

\smallskip
\noindent
\textbf{Results.} We now answer our research questions by comparing the runtime performance of \PyTRECEval{} to \TRECEval{} (\RQRef{1}) and a native Python implementation (\RQRef{2}).

\newcommand{\HDD}{HDD}
\newcommand{\SSD}{SSD}
\newcommand{\MMAP}{\texttt{tmpfs}}

\begin{figure*}[th!]
\centering
\newcommand{\inner}[2]{%
\begin{subfigure}[t]{0.33\textwidth}
\centering
\includegraphics[width=\textwidth]{resources/trec_eval_comparison/#1.pdf}
\caption{#2\label{fig:trec_eval:#1}}
\end{subfigure}}
\inner{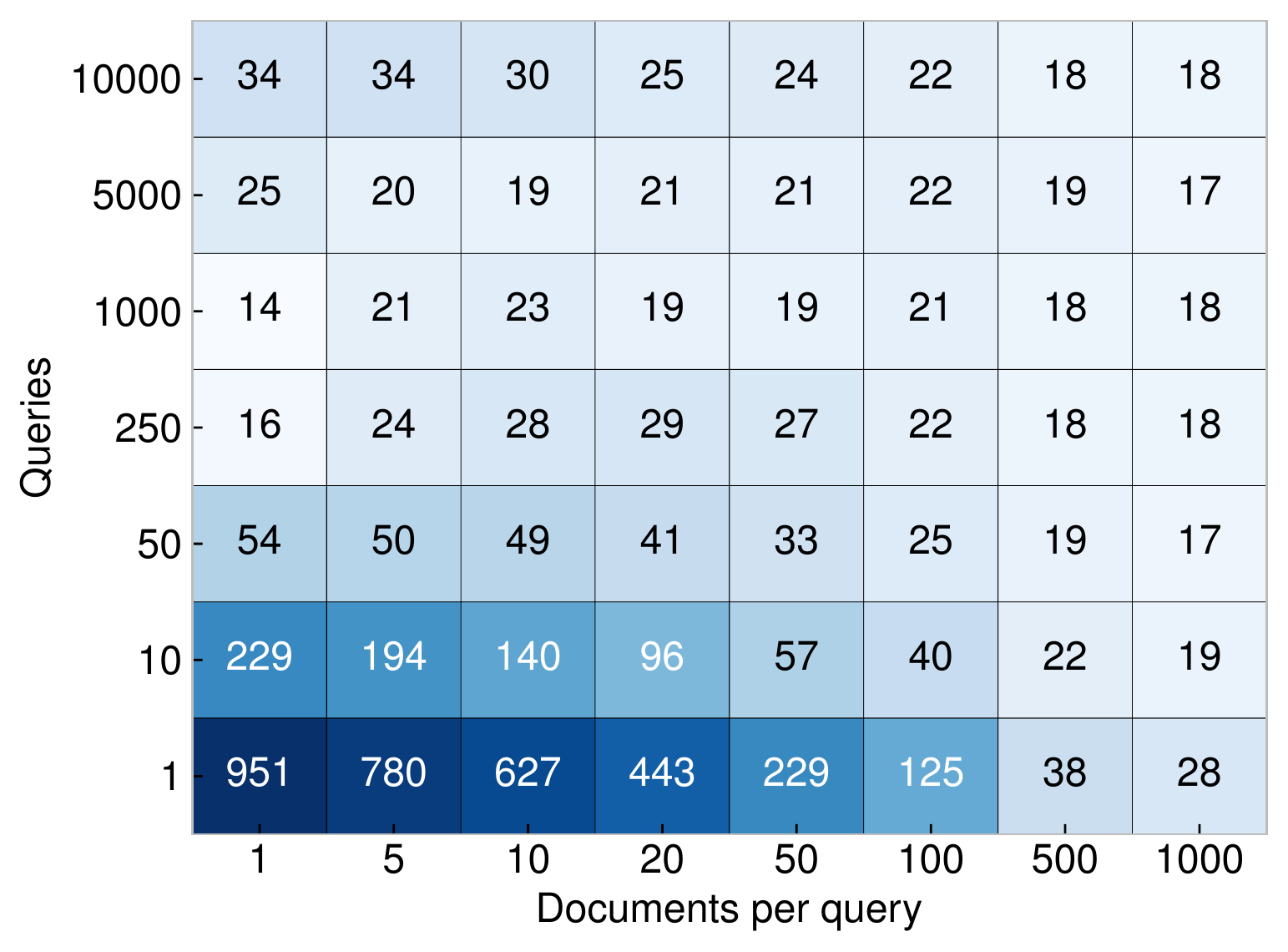}{HDD}
\inner{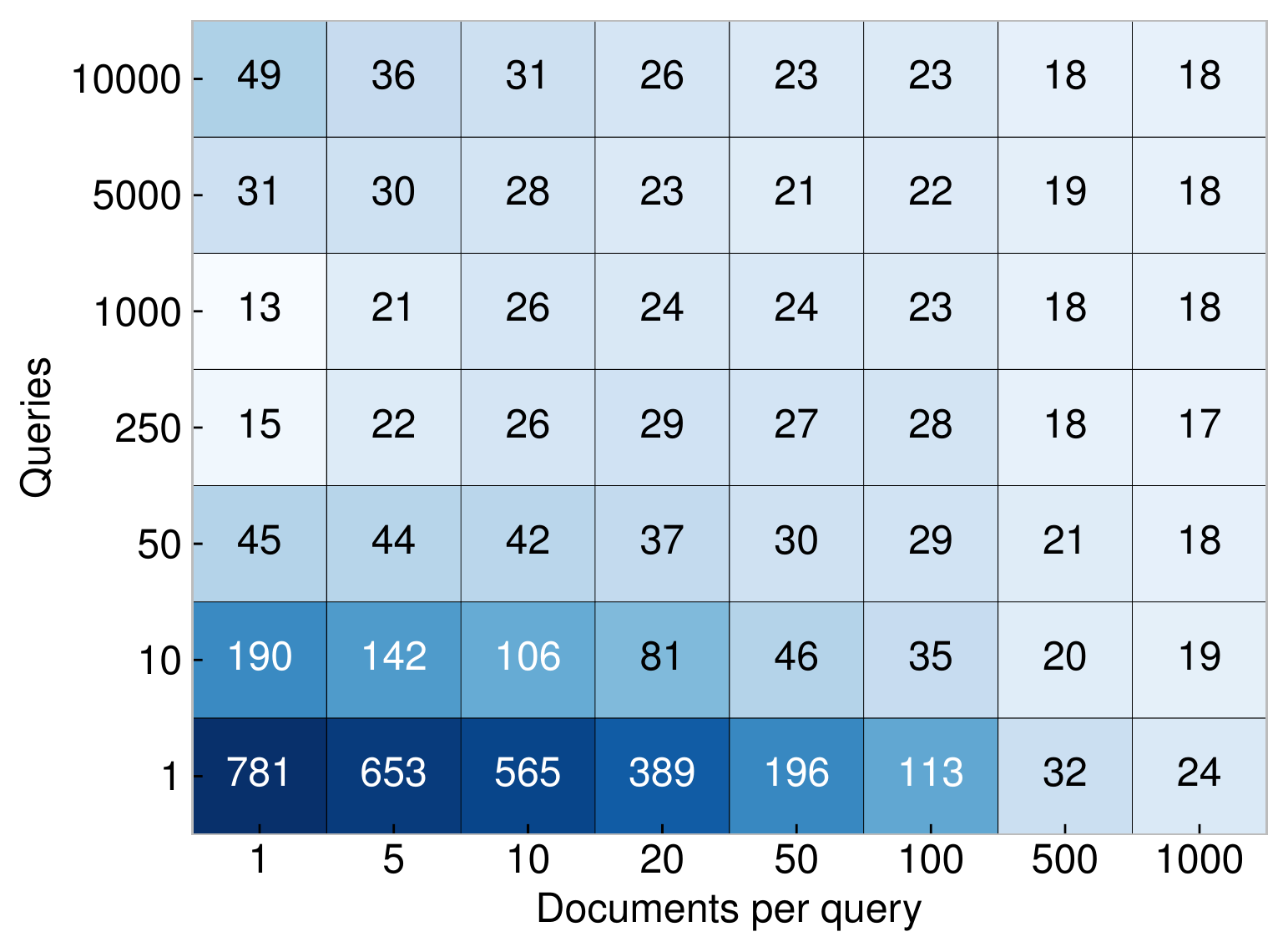}{SSD}
\inner{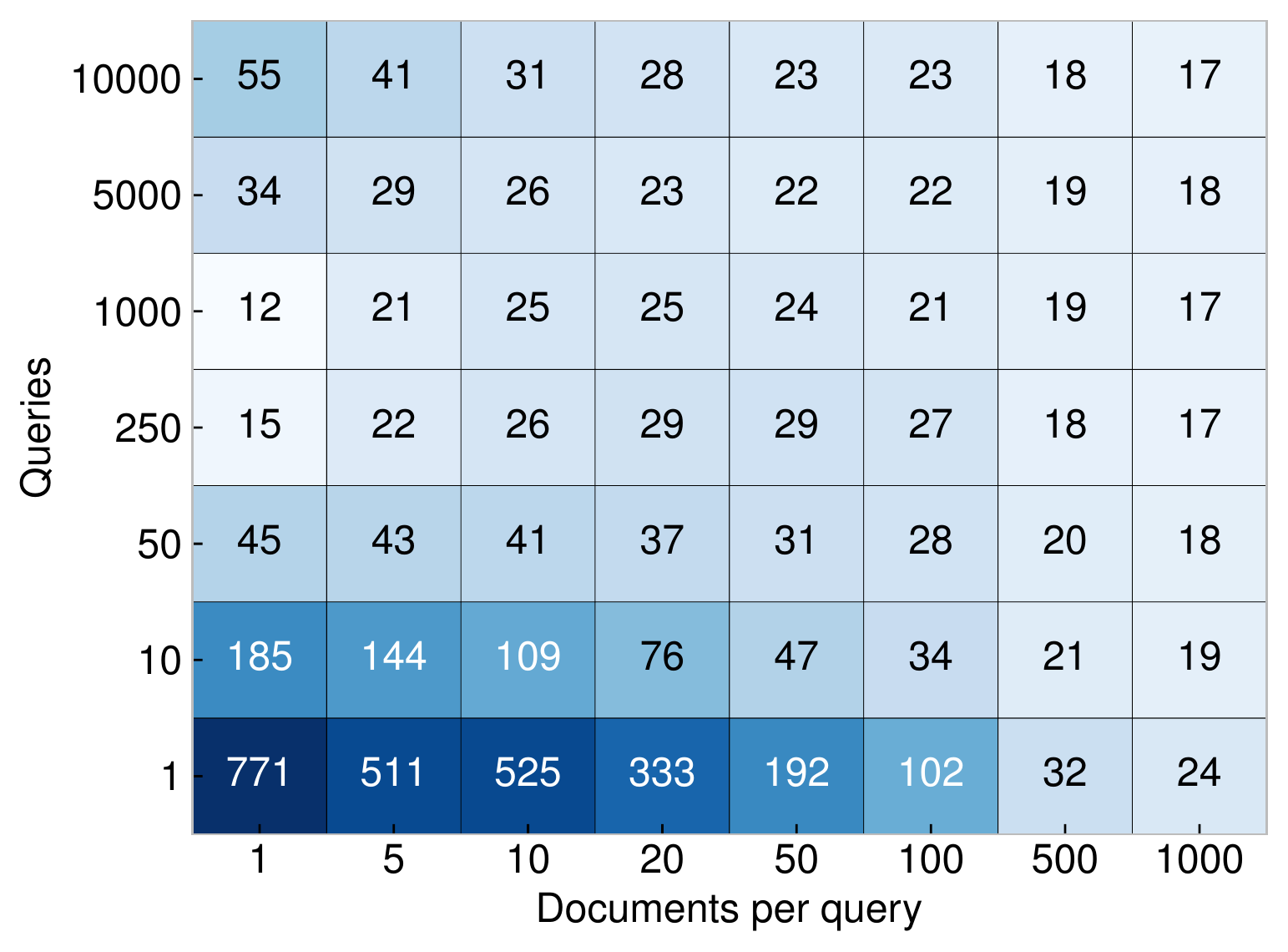}{Memory-mapped (\MMAP{})}
\caption{Speedup of \PyTRECEval{} (down-rounded speedup in each box; runtime measured as average over 20 repetitions) compared to invoking \TRECEval{} using a serialize-invoke-parse workflow (\S\ref{sec:introduction}) for different numbers of queries, different numbers of ranked documents per query, and using different types of storage (hard disk drive, solid state drive and random access memory) for serializing the rankings and query relevance ground truth.\label{fig:os}}
\end{figure*}

\RQAnswer{1}{\FirstRQ{}}{%
	Fig.~\ref{fig:os} shows matrices of speedups of \PyTRECEval{} over \TRECEval{} obtained using different storage types (increasing order of throughput capacity): a regular hard disk drive (\HDD{}), a solid state drive (\SSD{}) and a memory-mapped file system (\MMAP{}). For the degenerate case where we have a single query and a single returned document, we observe that there is a clear difference between the different storages. In particular, we can see that \MMAP{} is faster than \SSD{}, and in turn, \SSD{} is faster than the \HDD{}. However, for larger configurations (upper right box in every grid; \numprint{10000} queries with \numprint{1000} documents) we see that the difference between the storage types fades away and that \PyTRECEval{} always achieves a speedup of at least \numprint{17} over \TRECEval{}. This is because
	\begin{inparaenum}[(a)]
		\item starting the serialization (e.g., disk seek time) is expensive (as can be seen in the left-lower box of every grid), but that cost is quickly overshadowed by
		\item the cost of context switching between processes.
	\end{inparaenum}
	In the case of \PyTRECEval{}, however, context switching is avoided as all logic runs as part of the same process. Consequently, we can conclude that \PyTRECEval{} is at least one order of magnitude faster than invoking \TRECEval{} using a serialize-invoke-parse workflow.
}

\begin{figure}
\centering
\includegraphics[width=\columnwidth]{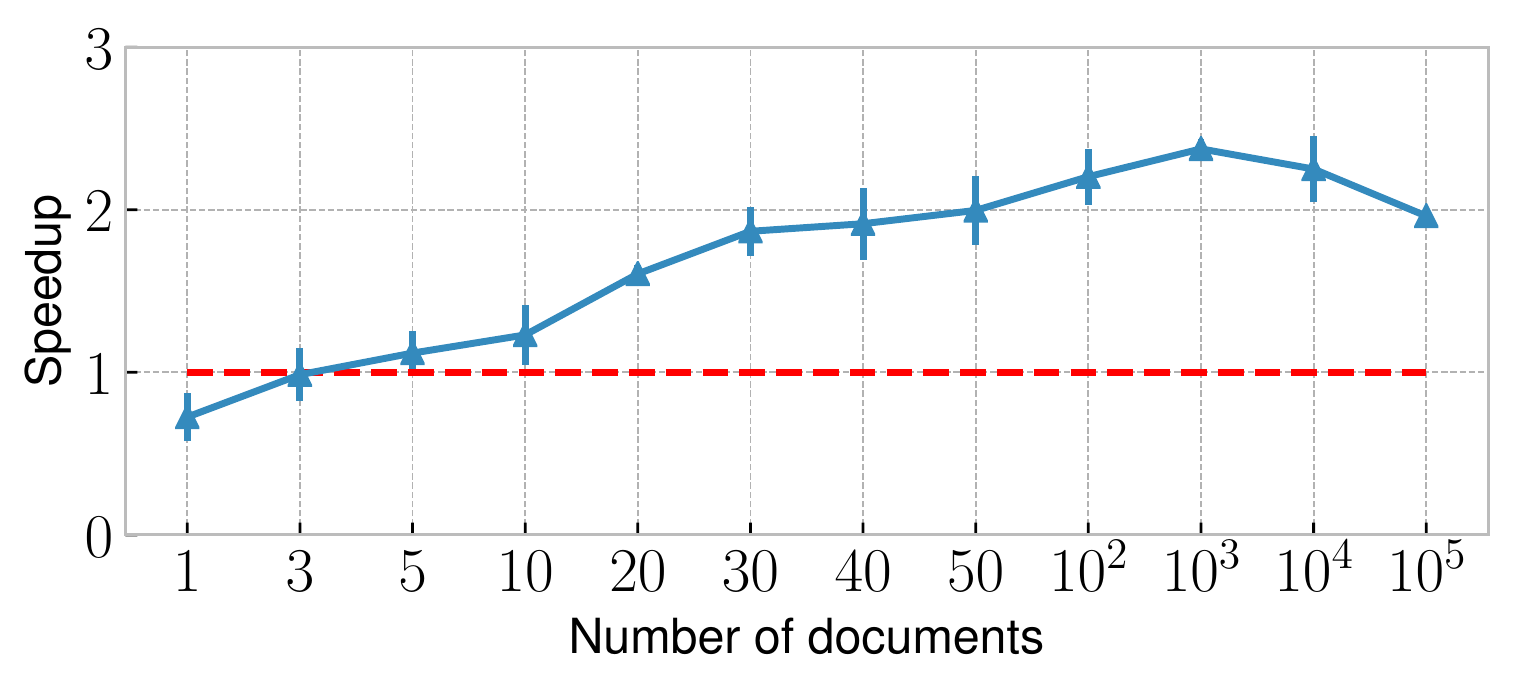}
\caption{Speedup of \PyTRECEval{} over a native Python implementation of the \NDCG{} evaluation measure (we report average speedup and its standard deviation over 20 repetitions). For practically-sized rankings, \PyTRECEval{} is consistently faster than the native Python implementation.\label{fig:native}}
\vspace*{-\baselineskip}
\end{figure}

\RQAnswer{2}{\SecondRQ{}}{%
	Fig.~\ref{fig:native} shows the speedup of \PyTRECEval{} over a Python-native implementation of \NDCG{} for a single query and a varying number of documents. Here we see that for extremely short rankings (1--3 documents), the native implementation outperforms \PyTRECEval{}. However, for rankings consisting of \numprint{5} documents or more, we can see that \PyTRECEval{} provides a consistent performance boost over the native implementation. The reason for the sub-native performance of \PyTRECEval{} for very short rankings is because---before \PyTRECEval{} computes evaluation measures---rankings need to be converted into the internal C format used by \TRECEval{}. The Python-native implementation does not require this transformation, and consequently, can thus be slightly faster when rankings are very short. However, it is important to note that short rankings are uncommon in IR and that the average ranking consists of around \numprint{100} to \numprint{1000} documents. We conclude that \PyTRECEval{} is faster than native Python implementations for practically-size rankings.
}

%% file: application.tex

\section{Example: Q-learning}

\begin{figure*}[t!]
\includegraphics[width=0.85\textwidth]{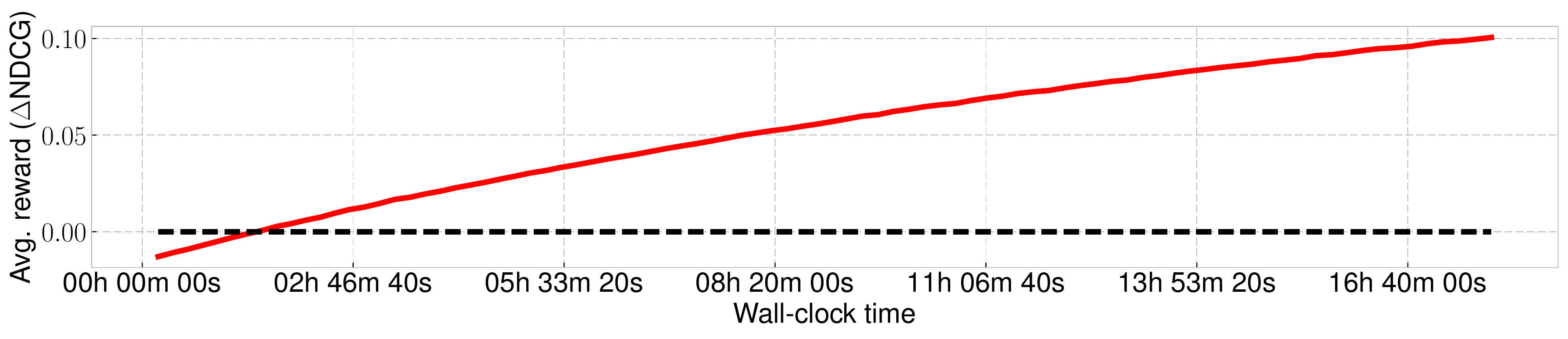}
\caption{Average reward ($\Delta$\NDCG{}) obtained by the Q-learning algorithm over time while training the reinforcement learning agent. The agent learns to select vocabulary terms that improve retrieval effectiveness for the set of 100k training queries.\label{fig:rl}}
\end{figure*}

We showcase the integration of the Pyndri indexing library \citep{VanGysel2017pyndri} and \PyTRECEval{} within the OpenAI Gym \citep{OpenAI2016}, a reinforcement learning library, for the task of query expansion. In particular, we use Pyndri to rank documents according to a textual query and subsequently evaluate the obtained ranking using \PyTRECEval{}. The reinforcement learning agent navigates an environment where actions correspond to adding a term to the query. Rewards are given by an increase or decrease in evaluation measure (i.e., \NDCG{}). The goal is for the agent to learn a policy $\pi^*$ that optimizes the expected value of the total reward. For the purpose of this demonstration of software interoperability, we synthesize a test collection in order to
\begin{inparaenum}[(1)]
	\item limit the computational complexity that arises from real-world collections, and
	\item to give us the ability to create an unlimited number of training queries and relevance judgments.
\end{inparaenum}

\newcommand{\Apply}[2]{#1\left(#2\right)}
\newcommand{\Prob}[1]{\Apply{P}{#1}}
\newcommand{\CondProb}[2]{\Prob{#1 \mid #2}}

\newcommand{\Length}[1]{\left|#1\right|}

\newcommand{\Vocabulary}{V}

\newcommand{\Documents}{D}
\newcommand{\Document}{\MakeLowercase{\Documents{}}}

\newcommand{\CorpusSize}{\Length{\Documents{}}}
\newcommand{\VocabularySize}{\Length{\Vocabulary{}}}
\newcommand{\AverageDocumentLength}{\mu_{\Document{}}}

\newcommand{\Queries}{Q}
\newcommand{\Query}{\MakeLowercase{\Queries{}}}

\newcommand{\AverageQueryLength}{\mu_{\Query{}}}

\newcommand{\NumRelevants}{r}
\newcommand{\QueryRelevants}{R_{\Query{}}}

\smallskip
\noindent
\textbf{Document collection.} We construct a synthetic document collection $\Documents{}$, of a given size $\CorpusSize{} = 100$, following the principles laid out by \citet{Tague1980pseudo}. For a given vocabulary size $\VocabularySize{} = \numprint{10 000}$, we construct vocabulary $\Vocabulary{}$ consisting of symbolic tokens. We sample collection-wide unigram ($\VocabularySize{}$ parameters) and bigram ($\VocabularySize{}^2$ parameters) pseudo counts from an exponential distribution ($\lambda = 1.0$). This incorporates term specificity within our synthetic collection, as only few term uni- and bigrams will be frequent and most will be infrequent. These pseudo counts will then serve as the concentration parameters of Dirichlet distributions from which we will sample a uni- and bigram language model for every document. We create $\CorpusSize{}$ documents as follows. For every document $\Document{}$, given the average document length $\AverageDocumentLength{} = 200$, we sample its document size, $\Length{\Document{}}$, from a Poisson with mean $\AverageDocumentLength{}$. We then sample two language models---one for unigrams $\CondProb{w}{\Document{}}$ and another for bigrams $\CondProb{\left(x, y\right)}{\Document{}}$---from a Dirichlet distribution where the concentration parameters we defined earlier for the whole collection. The document is then constructed as follows. Until we have reached $\Length{\Document{}}$ tokens, we repeat the following:
\begin{inparaenum}[(a)]
	\item sample an $n$-gram size from a predefined probability distribution ($\Prob{n = 1} = 0.9$, $\Prob{n = 2} = 0.1$), and subsequently,
	\item sample an $n$-gram from the corresponding language model.
\end{inparaenum}
We truncate a document if it exceeds its pre-defined length $\Length{\Document{}}$.

\smallskip
\noindent
\textbf{Query collection.} Once we obtained our synthetic document collection $\Documents{}$, we proceed by constructing our query set $\Queries{}$, of a given size $\Length{\Queries{}} = \numprint{100000}$, as follows. For every query $\Query{}$ to be constructed, we select $\NumRelevants{} = 5$ documents uniformly at random from $\Documents{}$ and denote these as the set of relevant documents $\QueryRelevants{} \subset \Documents{}$ for query $\Query{}$. Given the average query length $\AverageQueryLength{} = 3$, the length of query $\Query{}$, $\Length{\Query{}}$, is then sampled from a Poisson distribution with mean $\AverageQueryLength{}$. We write $\CondProb{w}{\QueryRelevants{}}$ and $\CondProb{w}{\Documents{}}$ to denote the empirical language models estimated from concatenating the relevant documents for query $\Query{}$ and from concatenating all documents in the collection $\Documents{}$ (i.e., the collection language model), respectively. The $\Length{\Query{}}$ terms of query $\Query{}$ are sampled with replacement from $\CondProb{w}{\QueryRelevants{}}\left(1.0 - \CondProb{w}{\Documents{}}\right)$, such that terms specific to $\QueryRelevants{}$ and uncommon in $\Documents{}$ are selected.

\smallskip
\noindent
\textbf{Environment.} For each query $\Query{}$, the environment is initialized to the state where only the query terms are present. At any given state, the agent can then choose to expand the query terms with any unigram term from the vocabulary $\Vocabulary{}$ in addition to a null operation action. Rankings are obtained by querying the Indri search engine using Pyndri, using a Dirichlet language model ($\mu = \numprint{2500}$), and obtaining a ranking of the top-10 documents. The reward of choosing an action is the $\Delta$\NDCG{} that is obtained by expanding the query with the chosen term. As observation, the agent receives a binary vector indicating which terms of the vocabulary $\Vocabulary{}$ occur at least once in the current expanded query. After \numprint{5} actions---or a perfect \NDCG{} (i.e., $1.0$) is achieved---the episode terminates.

\smallskip
\noindent
\textbf{Reinforcement learning agent.} We learn an optimal policy tabular $\pi^*$ using Q-learning where the initial values of the $\Apply{Q}{\cdot}$ are initialized to zero. We set the learning rate $\alpha = 0.1$ and the discount factor $\gamma = 0.95$. During learning, we maintain an $\epsilon$-greedy strategy with $\epsilon = 0.05$.
Fig.~\ref{fig:rl} shows the average reward obtained while training an agent on the reinforcement learning problem defined above. 
The average reward obtained by the agent increases over time. In particular, this example showcases that different IR libraries (Pyndri, \PyTRECEval{}) can easily be integrated with machine learning libraries (OpenAI Gym) to quickly prototype ideas. An essential part here is that expensive operations (i.e., ranking and evaluation) are performed in efficient low-level languages, whereas prototyping occurs in the high-level Python scripting language.
All code used in this example is available under the MIT open-source license.\footnote{The reinforcement learning code can be found in the \texttt{examples} sub-directory of the \PyTRECEval{} repository; see the footnote on the first page.}

%% file: conclusions.tex

\section{Conclusions}

In this paper we introduced \PyTRECEval{}, a Python interface to \TRECEval{}. \PyTRECEval{} builds upon the \TRECEval{} source code and exposes a Python-first interface to the \TRECEval{} evaluation toolkit as a native Python extension. This allows for convenient and fast invocation of IR evaluation measures directly from Python. We showed that \PyTRECEval{} is around one order of magnitude faster than invoking \TRECEval{} in a serialize-invoke-parse workflow as it avoids the costs associated with
\begin{inparaenum}[(1)]
	\item the serialization of the rankings to storage, and
	\item operation system context switching.
\end{inparaenum}
Compared to a native Python implementation of \NDCG{}, \PyTRECEval{} is approximately twice as fast for practically-sized rankings (\numprint{100} to \numprint{1000} documents). In addition, we showcased the integration of Pyndri \citep{VanGysel2017pyndri} and \PyTRECEval{} within the OpenAI Gym \citep{OpenAI2016} and showed that all three modules can be combined to quickly prototype ideas.

In this paper, we used a tabular function during Q-learning; other functional forms---such as a deep neural network---can also be used. 
Pyndri and \PyTRECEval{} expose common IR operations through a convenient Python interface. Beyond the convenience that both modules provide, an important design principle is that expensive operations (e.g., indexing, ranking) are performed using efficient low-level languages (e.g., C), while Python takes on the role of an instructor that links the expensive operations. Future work consists of exposing more IR operations as Python libraries and allowing more interoperability amongst modules. For example, currently Pyndri converts its internal Indri structures to Python structures, which are then again converted back to internal \TRECEval{} structures by \PyTRECEval{}. 
A closer integration of Pyndri and \PyTRECEval{} could result in even faster execution times as both can communicate directly---in cases where one is only interested in the evaluation measures and not the rankings---rather than through Python.